\documentclass[prl,preprint]{revtex4-1}
\usepackage{graphicx,amsmath}% Include figure files

\draft % marks overfull lines with a black rule on the right

\begin{document}

% Use the \preprint command to place your local institutional report number
% on the title page in preprint mode.
% Multiple \preprint commands are allowed.
%\preprint{}

\title{Coherent and incoherent electron-phonon coupling in graphite observed with radio-frequency compressed ultrafast electron diffraction} %Title of paper

% repeat the \author .. \affiliation  etc. as needed
% \email, \thanks, \homepage, \altaffiliation all apply to the current author.
% Explanatory text should go in the []'s,
% actual e-mail address or url should go in the {}'s for \email and \homepage.
% Please use the appropriate macro for the type of information

% \affiliation command applies to all authors since the last \affiliation command.
% The \affiliation command should follow the other information.

\author{Robert P. Chatelain}
\author{Vance R. Morrison}
\author{Bart L. M. Klarenaar}
\author{Bradley J. Siwick}
\email[Corresponding Author, email:  ]{bradley.siwick@mcgill.ca}
\affiliation{Departments of  Physics and Chemistry, Center for the Physics of Materials, McGill University, Montreal, Canada}

% Collaboration name, if desired (requires use of superscriptaddress option in \documentclass).
% \noaffiliation is required (may also be used with the \author command).
%\collaboration{}
%\noaffiliation

\date{\today}

\begin{abstract}
Radio-frequency compressed ultrafast electron diffraction has been used to probe the coherent and incoherent coupling of impulsive electronic excitation at 1.55 eV (800 nm) to optical and acoustic phonon modes directly from the perspective of the lattice degrees of freedom.  A bi-exponential suppression of diffracted intensity due to relaxation of the electronic system into incoherent phonons is observed, with the 250 fs fast contribution dominated by coupling to the $E_{2g2}$ optical phonon mode at the $\Gamma$-point ($\Gamma-E_{2g2}$) and $A_1^{'}$ optical phonon mode at the $K$-point ($K-A_1^{'}$).  Both modes have Kohn anomalies at these points in the Brillouin zone.  The result is a unique non-equilibrium state with the electron subsystem in thermal equilibrium with only a very small subset of the lattice degrees of freedom within 500 fs following photoexcitation.  This state relaxes through further electron-phonon and phonon-phonon pathways on the 6 ps timescale.  In addition, electronic excitation leads to both in-plane and out-of-plane coherent lattice responses in graphite whose character we are able to fully determine based on spot positions and intensity modulations in the femtosecond electron diffraction data.  The in-plane motion is specifically a $\Gamma$-point shearing mode of the graphene planes with an amplitude of approximately 0.06 pm and the out-of-plane motion an acoustic Ôbreathing modeÕ response of the film.

\end{abstract}

\maketitle %\maketitle must follow title, authors, abstract and \pacs
The unique electronic properties of graphite, graphene and carbon nanotubes have inspired a great deal of attention for both fundamental and practical reasons with the hope that these materials will enable high performance carbon-based electronic devices \cite{Allen2010,Bonaccorso2010}. In graphite and graphene, these properties emerge from the nature of the electron-lattice interactions in the highly anisotropic crystal structure made up of strong, covalently bound hexagonal sheets of carbon atoms (graphene) that are each weakly bound together to form the layered structure of graphite (Fig. 1b).  These interactions manifest most profoundly in the electronic bandstructure, notably the linear dispersion of the bands near the Fermi level and the point-like Fermi-surface at $K$ and $K^{'}$ in the Brillouin zone.  This linear dispersion results in the unique Ômass-lessÕ carrier transport properties and high electron mobilities for carriers near the Fermi level where the electron-phonon coupling is weak \cite{Leem2008}.  However, these interactions also manifest in other very important ways, in particular through the strongly anisotropic (momentum-dependent) electron-phonon interaction that results from the ineffective screening of the Coulomb interaction in these semimetals.  The most conspicuous manifestation of this coupling is the strong electronic renormalization in the optical phonon bands (Kohn anomalies) at the $K$ and $\Gamma$ points \cite{Piscanec2004}.  The strong nonadiabatic coupling between electrons and phonons for these optical $K - A^{Õ}_1$ and $\Gamma - E_{2g2}$ modes makes them profoundly sensitive to carrier doping \cite{Pisana2007} and the micro/nanostructure (e.g. number of layers, level of disorder) in a way that can be exploited through Raman spectroscopy as a materials characterization tool (through the so called D and G peaks that report on these modes) \cite{Ferrari2007}.  Interaction with these specific Ôstrongly coupled optical modesÕ (SCOP) has been implicated as the limiting factor in the high-field transport of carriers in graphene, graphite and carbon nanotubes \cite{Yao2000,Javey2004,Piscanec2004,Ferrari2007}.

The large body of work on this subject, both experimental and theoretical, speaks to the importance of understanding the electron-phonon interactions in graphene and graphite but also to its complexity.  Our work is unusual in that it focuses on the electron-phonon coupling from the perspective of the lattice degrees of freedom, rather than the electronic system or an electronic response function as has been more commonly the case.  Specifically, we probe the lattice responses to impulsive (35 fs, 800 nm)  photoexcitation using radio-frequency compressed ultrafast electron diffraction \cite{VanOudheusden2007,Chatelain2012,Gao2012}.  The 1.55 eV photons excite vertical $\pi-\pi^{*}$ transitions from the valence to conduction bands near the $K$ points to optically prepare a strongly non-equilibrium electron-hole plasma.  Previous experimental investigations of the photodoped, non-equilibrium carrier dynamics using time-resolved THz spectroscopy \cite{Kampfrath2005} and photoemission spectroscopy \cite{Ishida2011} have demonstrated that the initial carrier relaxation is extremely rapid, with approximately 90\% of the deposited electronic excitation energy leaving the electron sub-system within 500 fs.  The strongly coupled $K-A^{'}_1$ and $\Gamma-E_{2g2}$ modes were implicated as the reservoirs into which this energy preferentially flows, consistent with time-resolved Raman and vibrational spectroscopy studies of the $\Gamma-E_{2g2}$ mode \cite{Ishioka2008a,Yan2009}.  Benefitting here from the dramatic enhancements in both time-resolution and electron beam brightness that RF electron pulse compression affords, UED provides a complementary perspective on the electron-lattice coupling.  It enables us to directly probe both incoherent and coherent aspects of the lattice response in graphite following photoexcitation.  We are able to directly confirm that initial carrier relaxation is dominated by the interaction with the SCOP modes.  In addition, we observe coherent lattice responses beyond the incoherent coupling associated with the carrier relaxation/cooling, unifying a number of previous observations.

\begin{figure} [t]
 \centering
 \includegraphics[width = 1.0\textwidth]{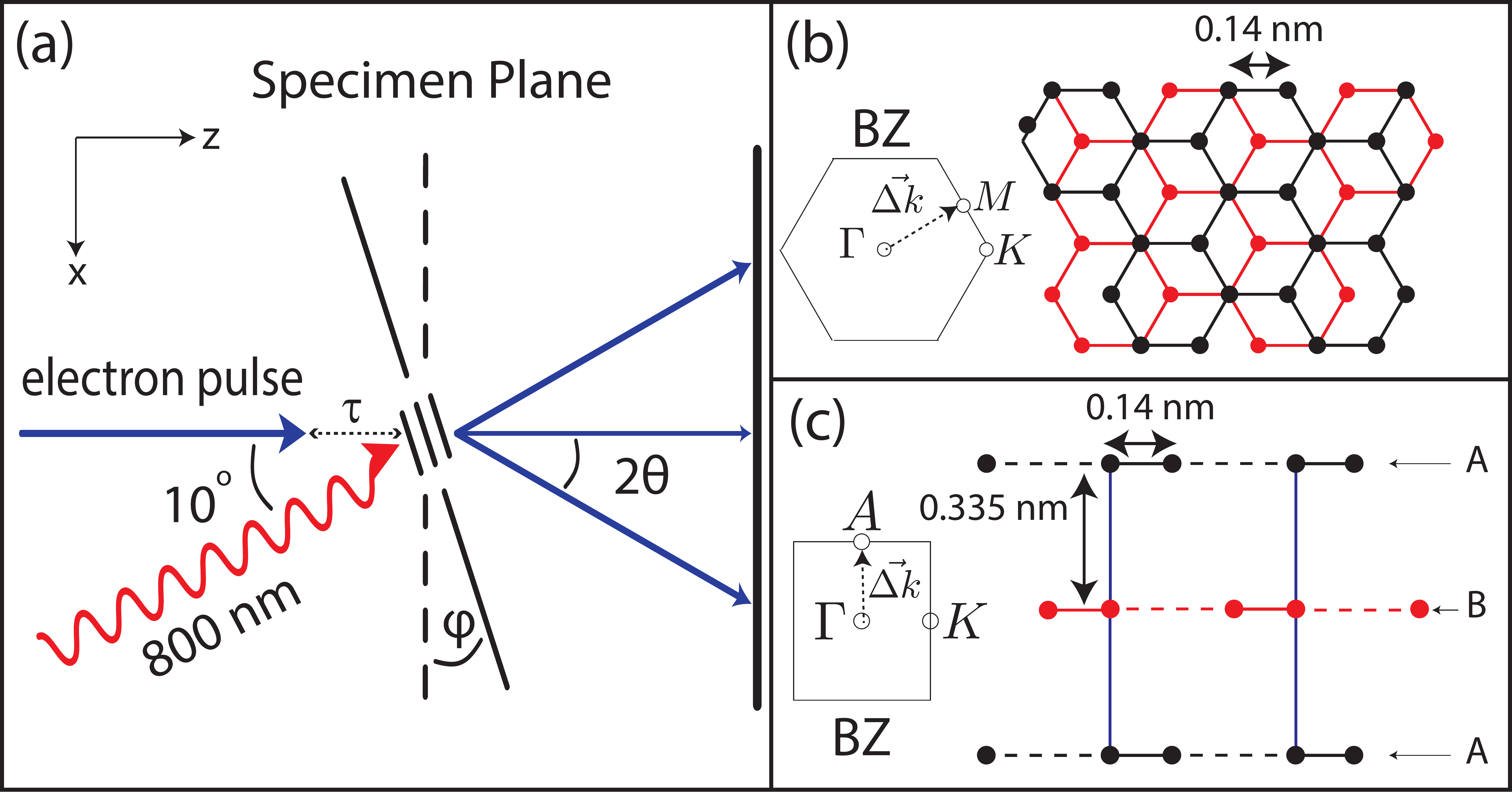}
 \caption{Ultrafast electron diffraction of single crystal graphite.  (a)  The specimen normal can be rotated by an angle $\phi$ relative to the transmitted electron beam direction.  Transient diffraction patterns are captured at every time delay $\tau$.  (b)  Unit cell and Brillouin zone of graphite in the x-y plane.  (c)  Unit cell and Brillouin zone of graphite in the x-z plane.}
 \label{fig1}
 \end{figure}

The single crystal specimens used in this work were obtained from Naturally Graphite$^{\copyright}$ and were repeatably exfoliated to thicknesses less than 50 nm using a technique similar to that introduced by Novoselov {\it et. al.} \cite{Novoselov2004}.   The specimens were mounted on a rotation ($\phi$) stage, allowing the interrogation of lattice dynamics in both the basal plane (graphene) and stacking direction (Fig. 1a). When $\phi = 0$ the electron beam direction is parallel to the surface normal and the $c$ axis of the graphite lattice.  Only Bragg reflections sensitive to in-plane structural dynamics ($h$$k$$l=0$) are observed in this orientation (Fig. 1b). For suitably chosen $\phi \neq 0$, however, reflections with $l\neq 0$ are observed and dynamics in both the stacking direction and graphene planes are probed (Fig. \ref{fig1}c).

\begin{figure} [t]
 \centering
 \includegraphics[width = 1.0\textwidth]{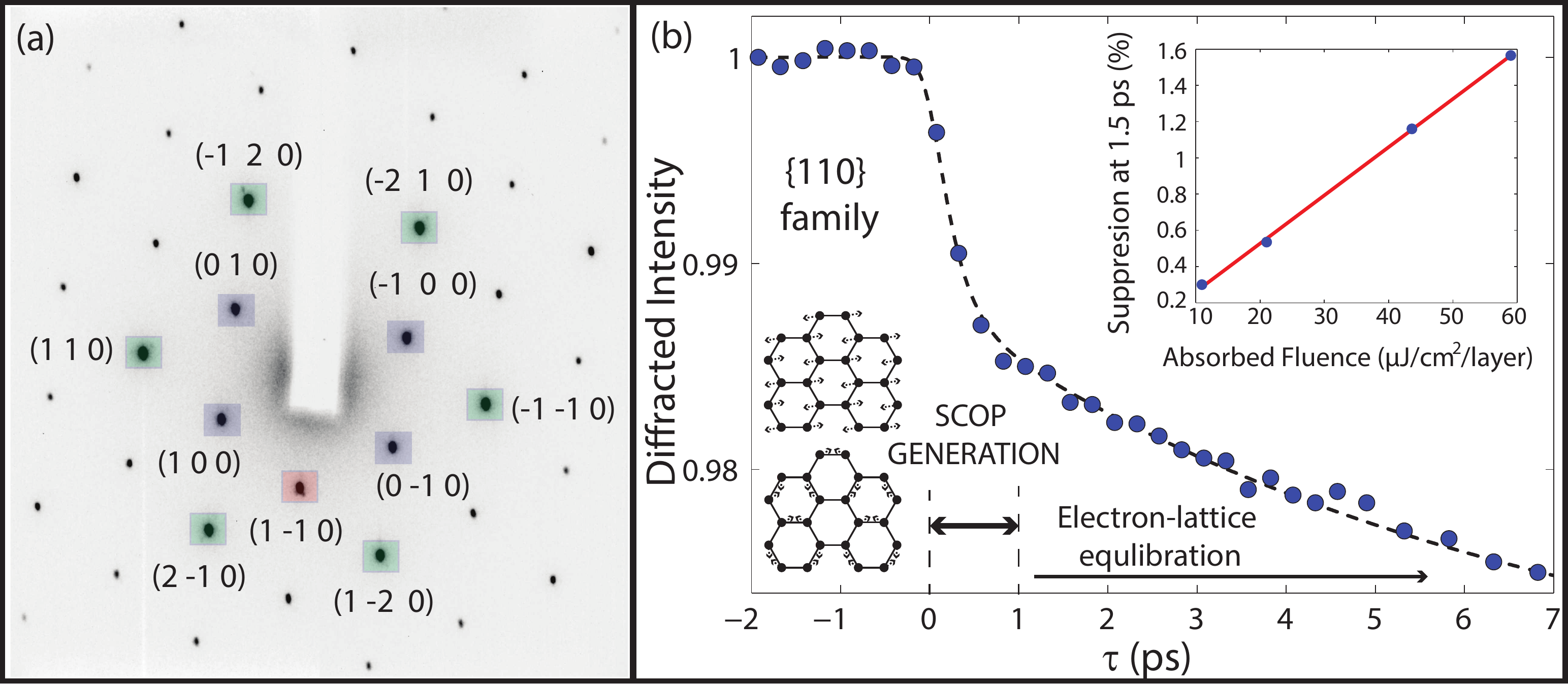}
\caption{Photoinduced lattice disorder of graphite measured in the \{110\} family of peaks.  (a)  Data obtained from peaks labelled by the same colored squares are averaged together.  (b)  Before time zero the lattice is at thermal equilibrium.  In the first picosecond following excitation, hot electrons couple to strongly coupled optical modes which in turn disorder the lattice.  At longer times, the sub-population of SCOP's thermalize with lower energy phonon modes as the lattice comes into thermal equilibrium with the electrons.}
 \label{fig2}
 \end{figure}

Experiments performed at normal incidence ($\phi=0$) reveal a bi-exponential behaviour of the \{110\} peak intensity dynamics.  Fig. \ref{fig2} shows data collected after laser excitation for a total absorbed fluence per layer of 59 $\mu$J/cm$^2$. The normalized intensity suppression is determined to a statistical precision better than 7 \textrm{x} $10^{-4}$ (approximately the size of the data markers) in 10 minutes of acquisition per time delay with RF compressed UED operating at 250 Hz \cite{Chatelain2012}.  The temporal instrument response function of the instrument has a width of $\sim$ 200 fs rms and the number of electrons per pulse is $\sim$ 0.1 pC. The dashed line is the result of fitting the function $I = I_0+A_1e^{-\tau/b_1}+A_2e^{-\tau/b_2}$ to the data.  The extracted time constants from the fitting procedure are 259 $\pm$ 60 fs and 6.5 $\pm$ 0.6 ps for $b_1$ and $b_2$, respectively, and the extracted normalized amplitudes of the suppressions are 0.013 $\pm$ 0.001 and 0.019 $\pm$ 0.001 for $A_1$ and $A_2$, respectively. An increase in dynamic disorder resulting from incoherent lattice excitation attenuates the zero Kelvin intensity of a diffraction peak by the Debye-Waller factor e$^{-2M}$, where
\begin{align}
M = 8\pi^2<u_s^2>\frac{\sin^2{\theta}}{\lambda^2},
\label{Meq}
\end{align}
and $<u_s^2>$ is the mean-square vibrational amplitude of atoms along the direction of the reciprocal lattice vector $\vec{g}_{hkl}$, where $|g_{hkl}|/2 = s= \sin{\theta}/\lambda$ \cite{Warren1969}. For incoherently populated modes, time averaging and two-dimensional isotropic averaging leads to the measured vibrational amplitude being related to the average vibrational amplitude, $a_P$, of any individual phonon by $<a_P^2> = 4<u_s^2>$. Classically, the vibrational energy per atom in a specific mode can also be related to the mode amplitude according to:
\begin{align}
\Delta <E_{vib}> = 2\pi^2 m \omega^2\Delta<a_P^2>,
\label{Ephonon}
\end{align}
where $m$ is the mass of a carbon atom and $\omega$ is the mode linear frequency.   Since the $K$ and $\Gamma$ modes are excited incoherently, increased population will result in an observed Debye-Waller like suppression of the peak intensities rather than a coherent modulation of diffracted intensity like that shown subsequently. If we assume that these two SCOP modes are the sole reservoirs into which electronic excitation energy flows on the $<$ 1 ps timescale, the extracted bi-exponential fit parameters (Fig. 2) given above, together with Eqs. 1 and 2 indicate that 89\% of the laser deposited energy is stored in the SCOP modes (where $\omega$ = 42.7 THz is the average SCOP frequency \cite{Maultzsch2004}) by $\tau$ = 500 fs. This result is in quantitative agreement with the rate of electronic energy relaxation reported earlier in photoemission \cite{Ishida2011} and THz studies \cite{Kampfrath2005}.  If lower-frequency modes played an appreciable role in the fast relaxation, the amplitude of the diffraction intensity suppression observed here could not be reconciled with the electronic energy relaxation observed previously.  For example, if the average frequency of the reservoir modes was 31 THz (approximately the average in graphite at equilibrium \cite{Nicklow1972}) the observed suppression would only account for 46\% of electronic excitation energy by 500 fs.  The bi-exponential behaviour of the UED data results from the closing of the SCOP relaxation channel once these specific modes reach thermal equilibrium with the electronic system  \cite{Butscher2007}, a process that also manifests in bi-exponential electronic energy relaxation \cite{Ishida2011} and related $\Gamma-E_{2g2}$ phonon frequency shifts \cite{Ishioka2008a,Yan2009}.   The longer timescale dynamics result in part from further carrier cooling via electron-phonon interactions with more weakly coupled mid-to-low energy phonon modes, but primarily the phonon-phonon scattering involved in lattice thermalization (i.e. cooling of the hot SCOP modes in which $\sim$89\% of the excitation energy is stored after 500 fs).

\begin{figure} [t]
 \centering
 \includegraphics[width = 1.0\textwidth]{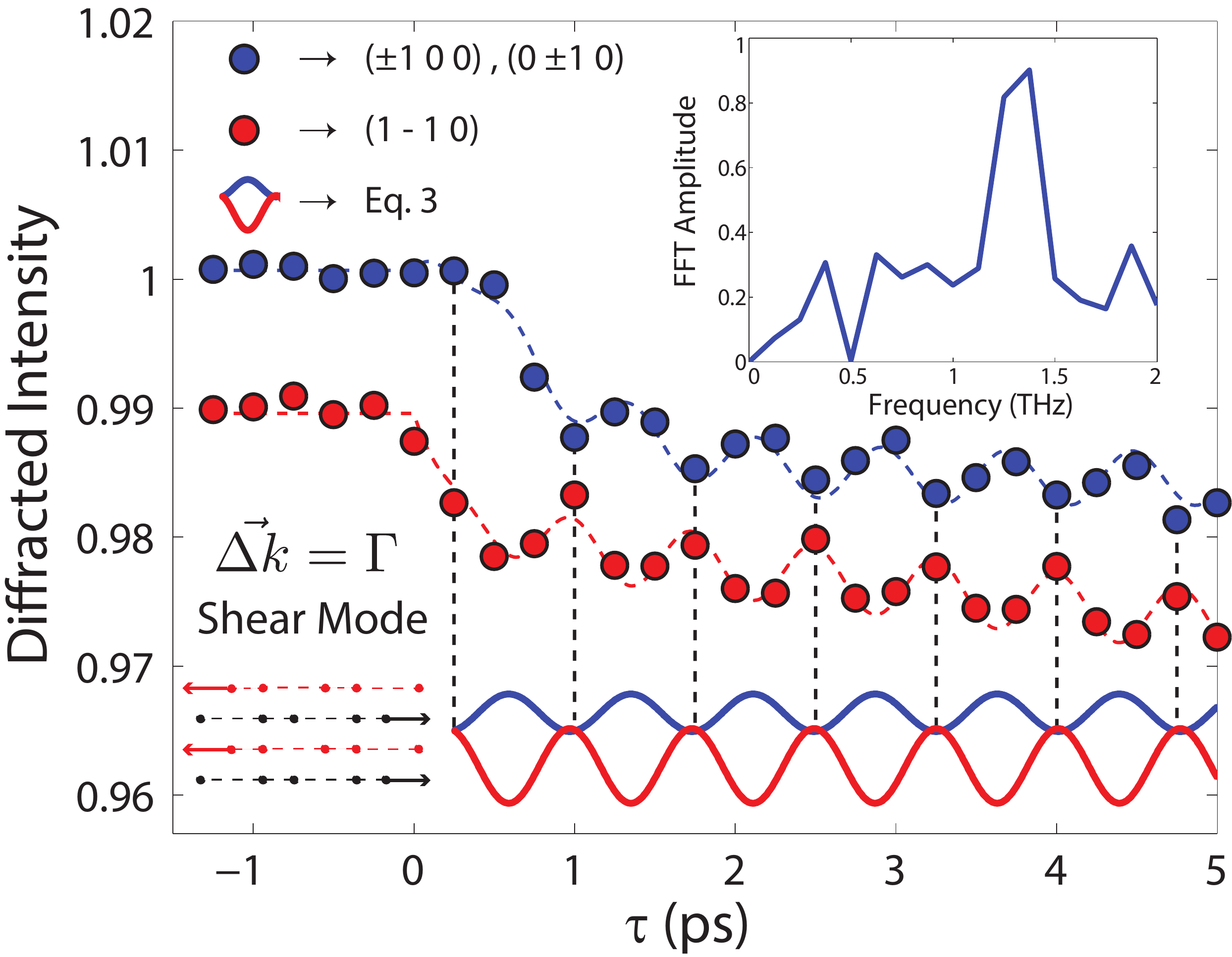}
 \caption{Coherent transverse shearing optical phonon of graphite.  As well as exhibiting peak suppressions, the \{100\} family of peaks have an oscillatory component to the intensity dynamics at early times.  The (1 -1 0) peak is out of phase by $\pi$ with respect to the other four measured peaks.  The optical phonon begins approximately 250 fs after the onset of the suppression dynamics and is well explained by equation \ref{structureFac}.}
 \label{fig3}
 \end{figure}

When phonons are excited coherently, the instantaneous lattice distortions that arise modulate the electron structure factor (and by extension the diffracted intensity) and/or the Bragg spot positions. Coherent coupling of the photoexcitation to in-plane and out-of-plane lattice motions were observed in addition to incoherent coupling to SCOP modes in these studies.  Fig. \ref{fig3} shows the peak intensity dynamics measured in the \{100\} family for the same dataset shown in Fig. \ref{fig2} (the blue and red data in Fig. \ref{fig3} was extracted from the corresponding peaks shown in Fig. \ref{fig2}a).  Superimposed on the intensity suppression is a modulation of intensity measured only in the \{100\} families.  Notice that the (1 -1 0) peak is out of phase with the remaining \{100\} peaks by precisely $\pi$.  The extracted modulation frequency---for both the red and blue data---is 1.3 THz  (see inset of Fig. 3).  This is in agreement with the measured frequency of the transverse optical shearing-mode phonon at the $\Gamma$-point \cite{Nicklow1972}.  This phonon affects peaks as a function of the indices $h$ and $k$ because the polarization vector of this mode points in the transverse direction.  Thus, experiments at $\phi=0$ measure these modes.  The anisotropy in the \{100\} peak dynamics was investigated by modelling the time-resolved structure factor of the $\Gamma$-point phonon:
\begin{align}
S(\tau) = 1+e^{-i2\pi\frac{2h+k}{3}}+e^{-i2\pi\frac{h-k}{3}A_f\sin \omega \tau}\left [1+e^{-i2\pi\frac{h+2k}{3}}\right],
\label{structureFac}
\end{align}
where $A_f$ is the shear amplitude and $\omega$ is the shear frequency.  The results of equation \ref{structureFac} for selected reflections are shown in Fig. \ref{fig3} as solid lines.  Equation \ref{structureFac} produces the correct out of phase relationship between the aforementioned peaks, while simultaneously predicting the lack of modulation of the \{110\} family (consistent with the data shown in Fig. \ref{fig2}).  For completeness, the $A$-point shear phonon was also modelled using an eight atom basis (rather than the four atom basis needed for the $\Gamma$-point mode).  The $A$-point mode does not explain the out of phase nature of the peak dynamics and suggests a much weaker modulation of peak intensities.    All measurements therefore suggest that a $\Gamma$-point shear mode along the [1 -1 0] real space direction was coherently excited.  The extracted shear amplitude from this analysis is 61 $\pm$ 19 fm.  Transient reflectivity measurements by Mishina {\it et. al.} suggest that the coherent excitation of the interlayer shearing mode is generated by the $\pi-\pi^{*}$ optical transition, and that the shear direction depends on the laser polarization \cite{Mishina2000}.  This directionality is directly observed here as an anisotropy in the diffraction peak intensity modulation. 

\begin{figure} [t]
 \centering
 \includegraphics[width = 1.0\textwidth]{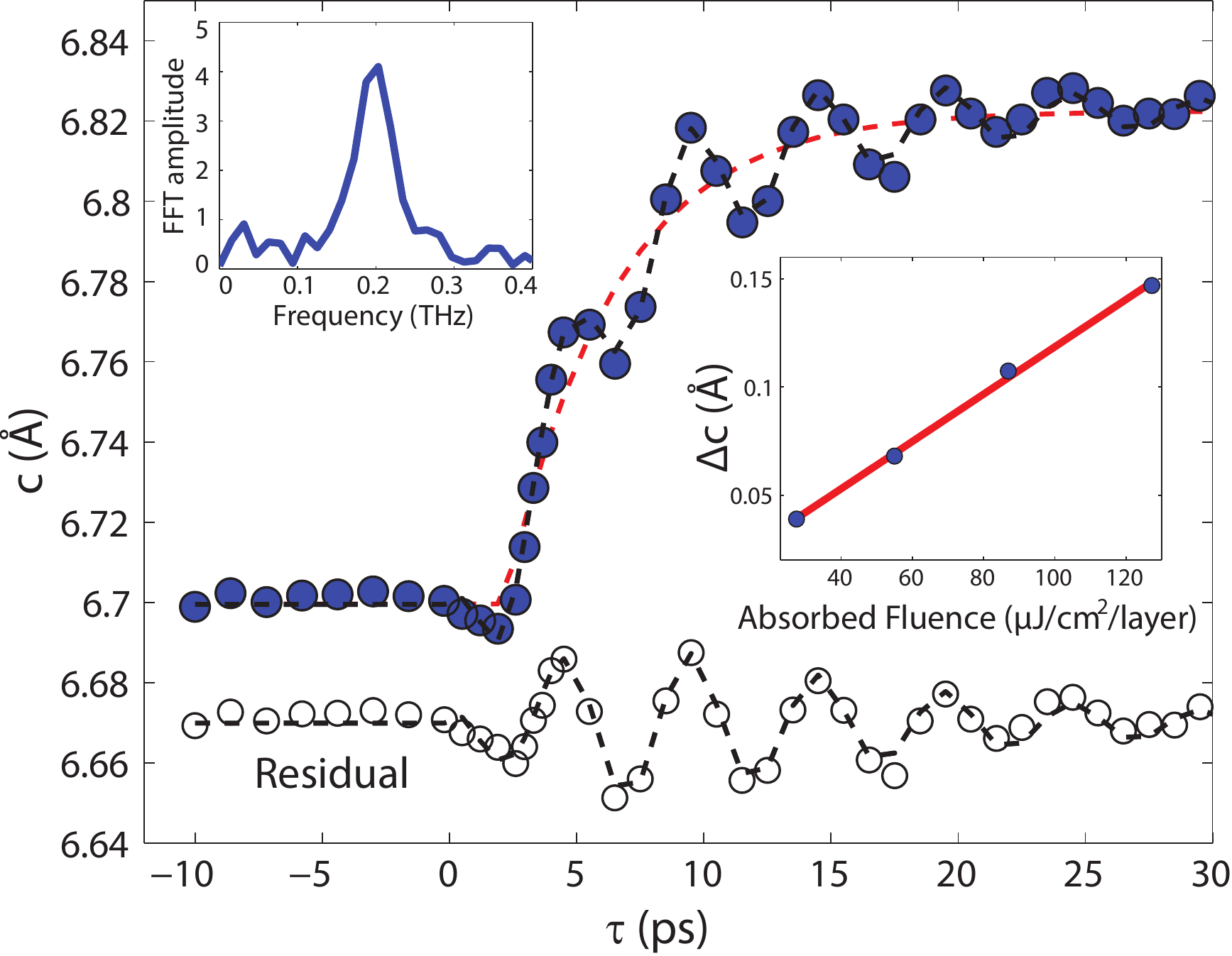}
 \caption{Lattice dynamics along the stacking direction.  The $c$ lattice constant exhibits an oscillatory behaviour during the photoinduced expansion that is best explained by the lowest order longitudinal acoustic phonon quantized by the sample thickness.  The measured frequency of 0.2 THz corresponds to a sample thickness of  10 nm.}
 \label{fig4}
 \end{figure}

Experiments performed at a larger angle ($\pi/8>\phi>\pi/4$) revealed statistically significant lattice expansion and coherent phonon signatures.  Coherent vibrations and lattice expansion dynamics have previously been observed in UED and UXD experiments \cite{Carbone2008,Raman2008,Carbone2009,Harb2011}.  Fig. \ref{fig4} shows the dynamics of the extracted $c$ lattice constant following excitation.
From experiments on different specimen thicknesses it was apparent that the frequency of the c-axis modulation depended strongly on the greyscale intensity (thickness) of the specimen when viewed through an optical microscope. The correlation between thickness and greyscale intensity was confirmed by AFM measurements.  The measured frequency is thus the lowest order longitudinal acoustic phonon mode quantized by the specimen thickness $L$.  Strong excitation of this mode is a feature of the dynamics in graphite specimens thinner than the optical absorption length ($\sim$140 nm).  Since the graphite is free standing, open boundary conditions were assumed and the specimen thickness is extracted using $L = v/2f$, where $v$ is the speed of sound and $f$ is the extracted linear frequency.  The vibration shown in Fig. \ref{fig4} is 0.2 THz and using the speed of sound of graphite in the stacking direction \cite{Bosak2007}, the specimen thickness is on the order of 10 nm.  The specimen used to collect the data shown in Figs. \ref{fig2} and \ref{fig3} had a frequency suggesting a thickness of 30 nm.

In conclusion, we have directly determined the incoherent and coherent electron-lattice couplings in graphite following femtosecond laser excitation using UED.  Within 500 fs, most of the initial electronic excitation energy has flowed out of the electron system into two specific SCOP modes, ie. $\Gamma-E_{2g2}$ and $K-A_1^{'}$.  The result is a unique non-equilibrium state with the electron subsystem in thermal equilibrium with only very a small subset of the lattice degrees of freedom. The coherent shear excitation also observed here demonstrates the sensitivity of UED when capturing subtle lattice distortions.  Not only can we extract vibrational periods, but we can localize the driven oscillation in $k$-space (to a specific wavenumber and direction) while simultaneously extracting vibrational amplitudes approximately seventy times the radius of a proton.  More generally, these results highlight the ability of UED to directly probe the fastest and most fundamental structural changes in material systems.

\section*{Acknowledgements}

This work was supported by the Canada Foundation for Innovation (CFI), Canada Research Chairs (CRC) program and the Natural Sciences and Engineering Research Council of Canada (NSERC). R.P.C. and V.M. gratefully acknowledges the support of NSERC PGS-D and CGS-D fellowships.  The authors also acknowledge fruitful discussions with Jom Luiten and his group.

%%%%%%%%% BIBLIOGRAPHY %%%%%%%%%%
%merlin.mbs apsrev4-1.bst 2010-07-25 4.21a (PWD, AO, DPC) hacked
%Control: key (0)
%Control: author (8) initials jnrlst
%Control: editor formatted (1) identically to author
%Control: production of article title (-1) disabled
%Control: page (0) single
%Control: year (1) truncated
%Control: production of eprint (0) enabled
%

%\bibliography{PRL.bib}
%%%%%%%%% END BIBLIOGRAPHY %%%%%%%%%%

%%%%%%%%% FIGURES %%%%%%%%%%

%% fig 1 %%%
 %%% fig 1 %%%

%%% fig 2 %%%
 %%% fig 2 %%%

 %%% fig 3 %%%
 %%% fig 3 %%%

  %%% fig 4 %%%
 %%% fig 4 %%%

 %%%%%%%%% END FIGURES %%%%%%%%%%

\end{document}